\documentclass[12pt]{article}

\usepackage{morefloats}
    \usepackage{amsmath}
    \usepackage{amsfonts}
\usepackage{color}
\usepackage{multirow}
\usepackage{graphicx}
\usepackage{rotating}

    \newcommand{\last}{\boldmath  {\large $\ast$} \unboldmath}

    \newtheorem{theo}{Theorem}

    \def\p{{\cal P}}

    \def\N{\mathbb{N}}
    \def\E{\mathbb{E}}
    
    \def\0{{\bf 0}}
    
    \def\R{\mathbb{R}}
    \def\PP{\mathbb{P}}

    \def\epsilon{\varepsilon}

    \def\Ty0M{{\cal T}_{y_0}\M}

    \renewcommand{\E}{\mathbb E \,}

    \newcommand{\tod}{\stackrel{{\cal D}}{\longrightarrow}}

    \newcommand{\eqco}{\setcounter{equation}{0}}
    \newcommand{\thco}{\setcounter{theo}{0}}
    \newcommand{\prco}{\setcounter{prop}{0}}
    \newcommand{\laco}{\setcounter{lemm}{0}}
    \newcommand{\coco}{\setcounter{coro}{0}}
    \newcommand{\cjco}{\setcounter{conj}{0}}
    
    \newcommand{\deco}{\setcounter{defn}{0}}
    
    \newcommand{\allco}{\eqco  \thco \prco \laco \coco \cjco \deco}
    \newcommand{\X}{{\cal X}}

\newcommand{\M}{{\cal M}}

    \renewcommand{\H}{{\cal H}}

    \newcommand{\Var}{{\rm Var}}

    \newcommand{\K}{{\cal K}}

    \def\bdm{\begin{displaymath}}
    \newcommand{\edm}{\end{displaymath}}
    \def\benu{\begin{enumerate}}
    \def\eenu{\end{enumerate}}
    \def\beqn{\begin{equation}}
    \def\eeqn{\end{equation}}
    \def\be{\begin{equation}}
    \def\ee{\end{equation}}
    \def\bea{\begin{eqnarray}}
    \def\eea{\end{eqnarray}}
    \newcommand{\bean}{\begin{eqnarray*}}
    \newcommand{\eean}{\end{eqnarray*}}
    \newcommand{\bear}{\begin{eqnarray}}
    \newcommand{\eear}{\end{eqnarray}}

    \def\R{\mathbb{R}}

    \def\qed{\hfill\hbox{${\vcenter{\vbox{
        \hrule height 0.4pt\hbox{\vrule width 0.4pt height 6pt
        \kern5pt\vrule width 0.4pt}\hrule height 0.4pt}}}$}}

\begin{document}

\title{\bf Multivariate goodness-of-fit on flat and curved spaces via nearest neighbor distances}

\author{ B. Ebner, N. Henze and J. E. Yukich$^{**}$}

\date{\today}
\maketitle

\footnotetext{ {\em American Mathematical Society 2000 subject
classifications.} Primary 62H15 Secondary 60F05, 60D05} 

\footnotetext{$~^{**}$ Research supported in part by NSF grant DMS-1406410}

\begin{abstract} We present a unified approach to goodness-of-fit testing in $\mathbb{R}^d$ and on lower-dimensional manifolds embedded in $\mathbb{R}^d$
based on sums of powers of weighted volumes of $k$-th nearest neighbor spheres. We prove asymptotic normality of a class of test statistics under the null
hypothesis and under fixed alternatives. Under such alternatives,  scaled versions of the test statistics converge to the $\alpha$-entropy
between probability distributions. A simulation study shows that the procedures are serious competitors to established goodness-of-fit tests.
\end{abstract}

\section{Introduction and summary}\label{secintro}
Nearest neighbor methods have been successfully applied in a variety of fields, such as classification (see \cite{Ga}),
density and regression function estimation (see \cite{Bi}, \cite{DV}), and
multivariate two-sample testing (see, e.g.,  \cite{He88}, \cite{Mo}, and \cite{Sc}). Moreover, nearest neighbor methods have also been employed in the context of testing the
goodness-of-fit of given data with a distributional model (see \cite{BiBr}, \cite{He82} and \cite{JSZ}).

This paper is devoted to a class of universally consistent goodness-of-fit tests based on nearest neighbors. These tests can be applied not only to test for uniformity on a compact domain
in $\mathbb{R}^d$, but also to test for a specified density on a $m$-dimensional manifold embedded in $\mathbb{R}^d$, where $m \le d$. Here, prominent special cases involve testing for uniformity on a circle or on a sphere.

To be specific, let $\M$ denote a $C^1$ $m$-dimensional manifold embedded in $\R^d$, where $m \leq d.$   $\M$
is endowed with the subset topology and is a closed subset of $\R^d$.
Let d$x$ be the Riemannian volume element
on $\M$.  A probability density function on $\M$ is a measurable non-negative real-valued function $f$ on $\M$ satisfying $\int_{\M} f(x)\, {\rm{d}}x = 1$.
The support $\K (f)$ of $f$ is the smallest closed set $K \subset \M$ such that $\int_K f(x) \, {\rm{d}} x =1$.

Let ${\p}(\M)$ denote the class of  bounded probability density functions
$f$ on $\M$, and write  ${\p}_b(\M) \subset \p(\M)$ for the subset of  probability density functions
$f$ such that  $\K (f)$ is compact and either (i) $\K (f)$ has no boundary or (ii) $\K (f)$ is a $C^1$
submanifold-with-boundary of $\M$; we refer to Section 2 of \cite{PY7} for details.
Notice that $\K (f)$ could be an $m$-sphere (or any ellipsoid) embedded in $\R^d$.
Let ${\p}_c(\M)$ denote those probability density functions $f \in
\p_b(\M)$  which are bounded away from zero on  their support.

In what follows we let $X_i$, $i \geq 1$,  be independent and identically distributed (i.i.d.) random variables with density $f$, defined on a common probability space $(\Omega,{\cal A},\mathbb{P})$,  and
we put $\X_n:= \{X_1,...,X_n\}$.

Given a locally finite subset $\X$ of $\M$ and $x \in \X$, we write $x^{(k)}$ for the $k$th nearest neighbor
(with respect to the Euclidean norm $| \cdot |$) of $x$ among $\X \setminus \{x \}$.  Let $v_m := \pi^{m/2}/ \Gamma(m/2 + 1)$
be the volume of the unit $m$-sphere.

Given a fixed $\alpha \in (0, \infty)$ and a fixed positive integer $J$, consider the volume score function induced by
the $J$ nearest neighbor distances:
\be \label{volscore}
\xi_{J}^{(\alpha)}(x, \X):= \sum_{k = 1}^J (v_m |x - x^{(k)}|^m)^{\alpha},
\ee
i.e., sums of volumes (to power $\alpha$) of the $k$ nearest neighbor balls around $x$, $k \in \{1,...,J\}$.  When $\X$ consists of
$\Theta(n)$ elements in a compact subset of $\M$, where $a n \le \Theta(n) \le b n$, $n \ge1$, for some $0<a<b<\infty$, we study the
re-scaled volume scores
\[ 
\xi_{n,J}^{(\alpha)}(x, \X):= \sum_{k = 1}^J (v_m |n^{1/m}(x - x^{(k)})|^m)^{\alpha}.
\] 
Recalling that $\X_n:= \{X_1,...,X_n\}$, we consider the random measure
\be \label{rmeasure}
\mu_{n,J}^{(\alpha)} := \sum_{X_i \in \X_n} \xi_{n,J}^{(\alpha)}(X_i, \X_n) \delta_{X_i},
\ee
with $\delta_x$ denoting the Dirac point mass at $x$.  If $h$ is an arbitrary measurable bounded function on $\M$, we write
$\langle \mu_{n,J}^{(\alpha)},  h \rangle$ for $\int_{\M} h(x)\, \textrm{d} \mu_{n,J}^{(\alpha)}(x)$.

Given a fixed $f_0 \in {\p}(\M)$, this paper considers testing goodness-of-fit of the hypothesis
\begin{equation}\label{hyph0}
H_0: {\textrm{ the unknown density of }} X_i \ {{\textrm is }} f_0,
\end{equation}
against general alternatives, based on the statistic
\be \label{statist}
T_{n,J}^{(\alpha)}:= \langle \mu_{n,J}^{(\alpha)},  f_0^{\alpha} \rangle := \sum_{X_i \in \X_n}   \xi_{n,J}^{(\alpha)}(X_i, \X_n) (f_0(X_i))^{\alpha}.
\ee
Notice that for the special case $m=d$  and $J=1$, this type of statistic has been studied in \cite{BiBr} and \cite{JSZ}, but without allowing for lower-dimensional manifolds, and without considering
fixed alternatives to $H_0$.

 In Section \ref{secmain}, we prove the asymptotic normality of $T_{n,J}^{(\alpha)}$ as $n$ tends to infinity both under $H_0$ and under fixed alternatives to $H_0$, and we show that $T_{n,J}^{(\alpha)}/n$ has an almost sure limit under a fixed alternative to
 $H_0$. In the case $0< \alpha < 1$, this limit is, apart from a multiplicative constant, the $\alpha$-entropy between $f$  and $f_0$. As a consequence, the statistic  $T_{n,J}^{(\alpha)}$ yields a universally goodness-of-fit test of $H_0$ for each $\alpha \in (0,\infty)$, $\alpha \neq 1$, and each $J$. The versatility of this class of tests is demonstrated in Section \ref{secsimul}, which presents the results of a simulation study comparing  our tests with several well-known competitors.  The paper concludes with some remarks and open problems.

\section{Main results}\label{secmain}

\allco
The limit theory for the statistic \eqref{statist} may be deduced from general theorems established in \cite{PY7} and goes as follows.

\begin{theo} \label{LLN} If $f \in {\p}_c(\M)$, $\alpha \in (0, \infty)$,  then as $n \to \infty$ we have
\be \label{LLN1}
\frac{T_{n,J}^{(\alpha)}}{n} \rightarrow  \sum_{k = 1}^J \frac{\Gamma( \alpha + k) } {\Gamma(k)} \int_{\M} f_0(x)^{\alpha}
f(x)^{1- \alpha} \, {\rm{d}}x
\ee
in $L^2$ and also ${\PP}$-a.s.
\end{theo}

\noindent{\em Remarks.}  (i)  Notice that the right-hand side of (\ref{LLN}) is distribution-free if $\alpha=1$. Thus, in view of the testing problem (\ref{hyph0}),
it is indispensable to have $\alpha \neq 1$.
 \vskip.3cm
(ii) If $\text{dim} \M = d$, if the support $\K(f)$ of $f$
 is a convex polyhedron, and if $f_0$ is the uniform density over $\M$, then the asserted $L^2$ convergence
   in \eqref{LLN1} is given by Theorem 2 of \cite{Wa}.  That paper, which is based on \cite{PY4}, shows that
\be \label{WadeLLN}
\E \xi_{n,J}^{(\alpha)}(X_1, \X_n) \rightarrow \int_{\K(f) } \E \xi_{J}^{(\alpha)}(\0, \H_{f(x)} ) \, f(x) \, {\rm{d}} x
\ee
holds in $L^2$ as $n \to \infty$. Here, $\0$ denotes a point at the origin
of $\R^m$, and $\H_\tau, \tau \in (0, \infty),$ stands for a homogeneous Poisson process of intensity $\tau$ in
$\R^m$, with $\R^m$ embedded in $\R^d$  so that the random variable
$\xi_{J}^{(\alpha)}(\0,\H_\tau)$ is well-defined.
 As will be shown in the upcoming proof, the paper \cite{PY7} upgrades \eqref{WadeLLN} to give convergence of the measures at \eqref{rmeasure}, it provides $L^2$ and
a.s. convergence, and also allows ${\cal K}(f)$ to be replaced by
a $C^1$ $m$-dimensional submanifold of $\R^d$.
\vskip.3cm
(iii) The statistic $T_{n,J}^{(\alpha)}$ may be considered a multivariate analogue of the statistic $\sum_{i=1}^{n+1} U_i^\alpha$ introduced in
\cite{Ki} for testing the hypothesis of a uniform distribution in the unit interval $[0,1]$. Here, $U_i= F(X_{(i)}) - F(X_{(i-1)})$, where
$0:= X_{(0)} \le X_{(1)} \le \ldots \le X_{(n)} \le X_{(n+1)} :=1$ are the order statistics of i.i.d. random variables $X_1,\ldots,X_n$
with common density function $f$. From the asymptotic distribution of $\sum_{i=1}^{n+1} U_i^\alpha$
(see, e.g., \cite{We}), it follows that
\[
\frac{1}{n} \sum_{i=1}^{n+1} (nU_i)^\alpha \rightarrow \Gamma(\alpha +1) \int_0^1 f(x)^{1-\alpha} \, \textrm{d}x
\]
in probability as $n \to \infty$. This result obviously corresponds to (\ref{LLN1})  for $J=1$ and $f_0$ being the uniform density over $\M$, where $m=d$ and $\M$ has Lebesgue measure one.
\vskip.3cm
(iv) If $\alpha \in (0,1)$, the integral
\[
r_\alpha(f_0,f) := \int_{\M} f_0(x)^{\alpha}
 f(x)^{1- \alpha} \, \textrm{d}x
\]
figuring on the right-hand side of \eqref{LLN1} is known as the $\alpha$-entropy between (the distributions associated with) $f_0$ and $f$,
see \cite{Va}.  Notice that $1-r_{1/2}(f_0,f) = H^2(f_0,f)$, where $H(f_0,f)$ is the Hellinger distance of $f_0$ and $f$.
By H\"older's inequality, $r_\alpha(f_0,f) \le 1$, with equality if and only if the distributions pertaining to $f_0$ and $f$ coincide.
If $\alpha \in (1,\infty)$, put $W:= f_0(X_1)/f(X_1)$, and recall that $X_1$ has density $f$. Then
\[
r_\alpha(f_0,f) = \E\left[ W^\alpha \right],
\]
and, by Jensen's inequality, $r_\alpha(f_0,f) \ge (\E W)^\alpha = 1$.
As above, equality holds if the distributions associated with $f_0$ and $f$ are the same.

\vskip.2cm
(v) It follows from (iii) and Theorem \ref{LLN} that, for fixed $\alpha \in (0,1)$, a test of fit that rejects the
hypothesis $H_0$ figuring in (\ref{hyph0}) for {\em small } values of $T_{n,J}^{(\alpha)}$ is consistent against each fixed alternative density
$f$. If $\alpha \in (1,\infty)$, rejection of $H_0$ is for {\em large}  values of $T_{n,J}^{(\alpha)}$, and the resulting test is universally consistent.

\vskip.3cm

Before stating variance asymptotics and a central limit theorem we introduce more notation from \cite{PY7}, especially (3.8) and (3.9) of that paper.   Given $u \in \R^m$, abbreviate $H_\tau \cup \{u\}$ by $H_\tau^u$.
We consider an integrated `covariance' of scores
\begin{align*}
 V(\tau):= V^{\xi_J^{(\alpha)}}(\tau) & := \E \xi_{J}^{(\alpha)}(\0, \H_\tau)^2 \\
  & \ \ \ \ +  \tau \int_{\R^m} \left\{ \E \xi_{J}^{(\alpha)}(\0, \H_\tau^{u})
\xi_{J}^{(\alpha)}(u, \H_\tau^{\0}) - (\E \xi_{J}^{(\alpha)}(\0,  \H_\tau))^2
\right\} \, \textrm{d}u
\end{align*}
and an integrated `add-one cost'
\[ 
\delta(\tau):=
 \delta^{ \xi_{J}^{(\alpha)} }(\tau) := \E \xi_{J}^{(\alpha)}(\0, \H_\tau) +  \tau \int_{\R^m} \E [\xi_{J}^{(\alpha)}( \0,  \H_\tau^{u}) -\xi_{J}^{(\alpha)}(\0, \H_\tau )] \, \textrm{d}u.
\] 
As shown in Theorem 3.2 of \cite{PY7}, these integrals are finite. Let N$(0, \sigma^2)$ denote a mean zero normal random variable with variance $\sigma^2$.

\begin{theo} \label{CLT} If $f \in {\p}_c(\M)$ is a.e. continuous and $\alpha \in (0, \infty)$,  then
\begin{align*} 
& \lim_{n \to \infty} n^{-1} \Var( T_{n,J}^{(\alpha)})  = \sigma^2(f_0, f) \\
&  :=  \int_{\M} f_0(x)^{2 \alpha} V(f(x)) f(x)\, {\rm{d}}x - \left( \int_{\cal M} \delta(f(x)) f_0(x)^{\alpha} f(x)\, {\rm{d}}x \right)^2 \in (0, \infty).
\end{align*}
Moreover, as $n \to \infty$,
\[ 
\frac{ T_{n,J}^{(\alpha)} - \E T_{n,J}^{(\alpha)}}
{ \sqrt{n}} \tod {\rm{N}}(0, \sigma^2(f_0, f)).
\] 
\end{theo}

\vskip.5cm

\noindent{\em Remark.} Theorem 2.1 of \cite{BPY} provides variance asymptotics and a central limit theorem for sums of functions of $k$th nearest neighbor distances in the special case $m=d$.
\vskip.5cm

\noindent {\em{Proof of Theorem \ref{LLN}}}.  We deduce this from Theorem 3.1 of \cite{PY7} with $\rho = \infty$, especially display (3.16) of \cite{PY7}, with
the $f$ in (3.16) of \cite{PY7} set to $f_0^{\alpha}$ and with the $\kappa$ in (3.16) of \cite{PY7} set to $f$.  Observe that $\xi_J^{(\alpha)}$ belongs to the class $\Sigma(k, r)$ of that paper, and notice that
$$
\sup_n \E ( \xi_{n,J}^{(\alpha)}(X_1, \X_n) )^p < \infty
$$
holds for all $p \in [1, \infty)$, i.e., the moment condition (3.4) of \cite{PY7} holds for all $p$. 

The limit (3.16) of \cite{PY7} tells us that as $n \to \infty$ we have convergence in $L^2$
\be \label{2LLN}
\frac{T_{n,J}^{(\alpha)}}{n}  \to \int_{\M} f_0(x)^\alpha \E \xi_J^{(\alpha)}(\0, \H_{f(x)}) f(x) \, \textrm{d}x,
\ee
where $\xi_J^{(\alpha)}(\0, \H_{f(x)})$ is defined at \eqref{volscore}. The last assertion in Theorem 3.1 of \cite{PY7} also gives a.s. convergence in (\ref{2LLN}).

Given $\tau \in (0, \infty)$ and  $\H_{\tau}$, we let $X_\tau^{(k)} \in \H_{\tau}$ be the $k$th nearest neighbor to the origin.  We compute
\begin{align*}
 \E \xi_J^{(\alpha)}(\0, \H_{\tau})
& =  \sum_{k = 1}^J \E (v_m |X_\tau^{(k)}|^m)^{\alpha} \\
& =  \sum_{k = 1}^J  v_m^{\alpha}  (\tau^{-1/m} )^{\alpha m}  \E ( |X_1^{(k)}|)^{\alpha m} \\
& =   \left(\frac{v_m}{\tau}\right)^{\alpha} \sum_{k = 1}^J  v_m^{- (\alpha m)/m } \frac{ \Gamma(k + \alpha) } {\Gamma(k) } \\
& = \tau^{-\alpha}  \sum_{k = 1}^J \frac{ \Gamma(k + \alpha) } {\Gamma(k) },
\end{align*}
where the penultimate equality follows by display (15) of \cite{Wa} (with $\alpha$ replaced by $\alpha m$, $d$ replaced by $m$). 
 We have thus shown
\be \label{3LLN}
\E \xi_J^{(\alpha)}(\0, \H_{\tau}) =  \tau^{-\alpha}  \sum_{k = 1}^J \frac{ \Gamma(k + \alpha) } {\Gamma(k) }.
\ee
Letting $\tau$  equal $f(x)$ in \eqref{2LLN} and applying  \eqref{3LLN} gives the claimed limit \eqref{LLN1}.   \qed

\vskip.5cm
\noindent {\em{Proof of Theorem \ref{CLT}}}. This is an immediate consequence of Theorem 3.2 of \cite{PY7} as well as remark (iv)
on p. 2174 of \cite{PY7}.  In that remark we may set the function $f$ there to $f_0^\alpha$, we set $\rho = \infty$, and we put
$\mu_{n, k, \rho}^\xi$ equal to $\mu_{n,J}^{(\alpha)}$.  Keeping $\rho$ set to infinity, it is a straightforward matter to show that $\mu_{n,J}^{(\alpha)}$ satisfies the
moment conditions (3.5) and (3.6) of \cite{PY7}. 
 Since  $\mu_{n,J}^{(\alpha)}$ satisfies all the conditions
of remark (iv) on p. 2174 of \cite{PY7}, Theorem \ref{CLT} follows as desired.   \qed

\section{Simulations}\label{secsimul}
By means of a simulation study, this section compares the finite-sample power performance of the test based on $T_{n,J}^{(\alpha)}$ with that of several competitors. All simulations are performed using the statistical computing environment {\tt R}, see \cite{rco:16}. We consider testing for uniformity on the unit square $[0,1]^2$, on the unit circle $\mathcal{S}^1 = \{x \in \mathbb{R}^2: |x|=1\}$, and on the unit sphere
$\mathcal{S}^2 = \{x \in \mathbb{R}^3: |x | =1\}$. Since, strictly speaking, there is not only one new test, but a whole family of tests that depend on the choice of the power $\alpha$ and the number $J$ of neighbors taken into account, the impact on finite-sample power of $\alpha$ and $J$ will be of particular interest.  In each scenario, we consider the sample sizes $n=50$, $n=100$ and $n=200$, and the nominal level of significance is set to $0.05$. Throughout, critical values for $T_{n,J}^{(\alpha)}$ under $H_0$ (see Tables \ref{crit1} -- \ref{crit3}) have been simulated with $100~000$ replications, and each entry in a table referring to the power of the test
is based on $10~000$ replications.

\subsection{Unit square $[0,1]^2$}
For testing the hypothesis $H_0$ that the distribution of $X_1$ is uniform over the unit square $[0,1]^2$,
we considered the following competitors to the new test statistic.
\begin{itemize}
\item [(i)] The Distance to Boundary Test $DB$ (see \cite{BCV}), which is based on the distance of $X_1,\ldots,X_n$ to the boundary $\partial W$ of  $W:=[0,1]^2$.
Writing $D_B(y,\partial W):=\min\{|x-y|: x\in\partial W\}$ for the distance of $y \in W$ to $\partial W$ and $R:=\max\{D_B(x,\partial W):x\in W\}$  for the largest of such distances (which equals $0.5$
 in our case), the test statistic computes the values
    \begin{equation*}
    Y_j:=\frac{D_B(X_j,\partial W)}{R},\quad j=1,\ldots,n.
    \end{equation*}
       Under $H_0$ the random variables $Y_1,\ldots,Y_n$ have a  $\mbox{Beta}(1,2)$-distribution. The test employs the Kolmogorov-Smirnov type statistic
    \begin{equation*}
    DB_n=\sqrt{n}\sup_{y\in[0,1]}|G_n(y)-G_0(y)|.
    \end{equation*}
    Here, $G_n$ is the empirical distribution function of $Y_1,\ldots,Y_n$, and $G_0$ is the distribution function of the $\mbox{Beta}(1,2)$-distribution. Rejection of $H_0$ is for large values of $DB_n$, and critical values can be taken from the Kolmogorov distribution. Note that this test is not consistent against some easily computable alternatives, e.g., the uniform distribution on the subset $[0.5,1]^2$ of $W$.

\item[(ii)] The Maximal Spacing Test $MS$, see \cite{BCP}. Writing $B(x,r)$ for an open circle centered at $x$ with radius $r$, this test considers the maximum radius
 \begin{equation*}
  \Delta_n:=\sup\{r>0: \mbox{ there is some } x\,\mbox{with}\,B(x,r)\subset [0,1]^2\setminus \X_n\}
 \end{equation*}
 of a circle that does not contain  any of $X_1,\ldots,X_n$ as an inner point. Rejection of $H_0$ is for large values of the test statistic $V_n :=\pi \Delta_n^2$. The limit distribution of $V_n$ under $H_0$ follows from
 (2.5) of \cite{Ja87}, which states that
 \begin{equation*}
  nV_n-\log n - \log \log n \tod G,\quad n\rightarrow\infty,
 \end{equation*}
 where the random variable $G$ follows a Gumbel distribution with distribution function $\exp(-\exp(-x))$, $x \in \mathbb{R}$. Letting $u_\alpha$ denote the
  ($1-\alpha$)-quantile of this distribution, the test rejects $H_0$ at asymptotic level $\alpha$ if
  \begin{equation*}
  V_n>n^{-1}(u_\alpha+\log n + \log \log n).
 \end{equation*}
\end{itemize}
Nothing is known regarding the consistency of this test.

Since dealing with nearest neighbors in the square involves boundary effects (see, e.g., \cite{DH}), we initially employed  both the Euclidean metric and the
torus metric, i.e.,  the Euclidean metric on the 3d-torus, obained as the quotient of the unit square by pasting opposite edges together
via the identifications $(x,y) \sim (x+1,y) \sim (x,y+1)$. Because the power of the tests was in general somewhat higher for
the torus metric than for the Euclidean metric, we decided to use the torus metric. It should be stressed that this choice conforms to the general set-up
adopted in \cite{PY7} so that Theorem \ref{LLN} and Theorem \ref{CLT} remain valid.

An empirical study of uniformity tests in several settings including the hypercube can be found in \cite{Pet}. Guided by the simulation study in \cite{BCP2}, we used a contamination and a clustering model as alternatives
to the uniform distribution. The contamination model, denoted by CON,  for the distribution of $X_1$ is the mixture 
\begin{equation*}
(1-\varepsilon_1-\varepsilon_2){\rm U}[0,1]^2+\varepsilon_1 {\rm N}_2(c_1,\sigma_1^2)+\varepsilon_2{\rm  N}_2(c_2,\sigma_2^2),
\end{equation*}
conditionally on $X_1 \in [0,1]^2$. Here, $\varepsilon_1=0.135$, $\varepsilon_2=0.24$, $\sigma_1=0.09$, $\sigma_2=0.12$, $c_1= (0.25,0.25)$, $c_2 = (0.7,0.7)$, ${\rm U}[0,1]^2$ is the uniform
distribution over $[0,1]^2$,  and ${\rm N}_2(c_j,\sigma_j^2)$ stands for the bivariate normal distribution
with expectation vector $c_j$ and covariance matrix $\sigma_j^2 {\rm I}_2$, where ${\rm I}_2$ is the identity matrix of order 2. In other words, this model produces a uniform background noise and two radially symmetric point sources of data, centered at the points $c_1$ and $c_2$. The additional specification "conditionally on $X_1 \in [0,1]^2$" means that a realization was discarded whenever the generated point did not fall into the unit square.

\begin{center}
\begin{figure*}
  \includegraphics[width=7cm, height=7.25cm]{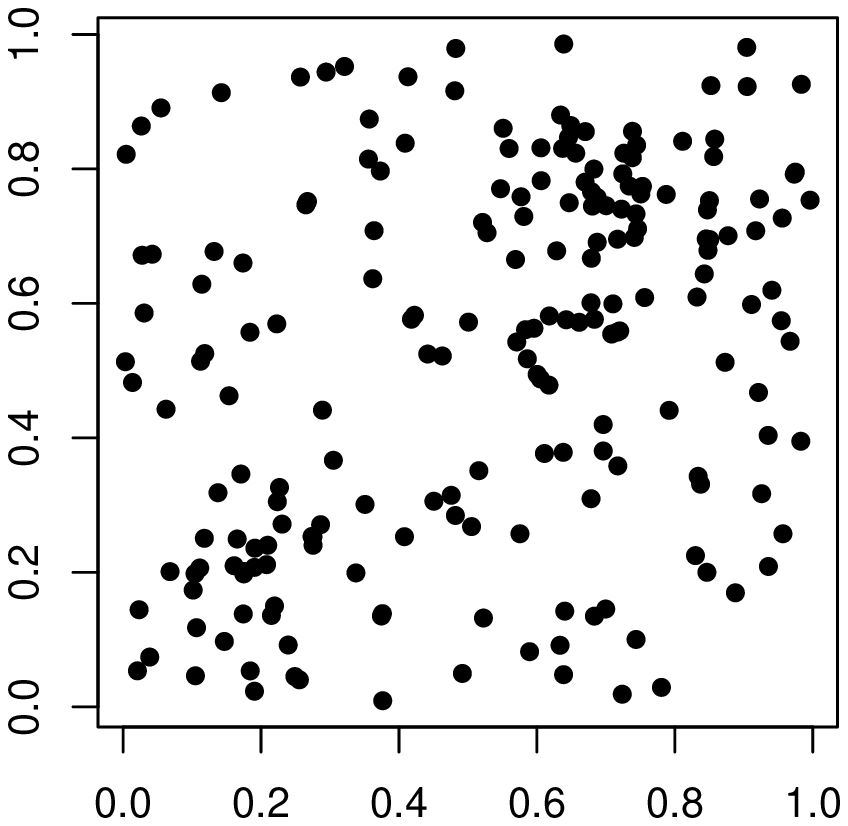}\hspace{0.5cm}\includegraphics[width=7cm, height=7.25cm]{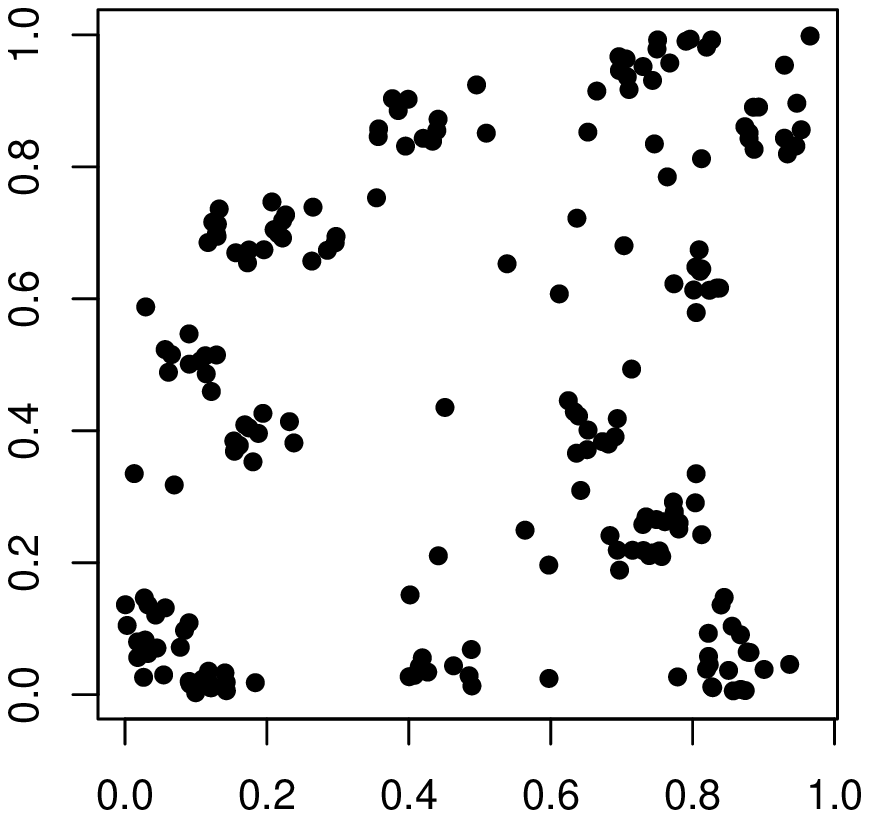}
\vspace*{-5mm}
\caption{Realization of the CON model (left) and the CLU model (right), $n=200$}\label{CONTCLU}     
\end{figure*}
\end{center}

The clustering alternative CLU (say) considers an alternative to $H_0$ in the non i.i.d. case, using a two step-technique. In a first step, one simulates $n_1=10$ i.i.d. random points with the uniform distribution
U$[0,1]^2$, which are then discarded but play the role of centers of clusters. In a second step, one generates, independently of each other, for each of those $n_1$ centers
$n_2=n/n_1$ points that are uniformly distributed in a disc of radius $0.05$,
the midpoint being the center. Similar to the CON alternative, each point was discarded if it fell outside $[0,1]^2$, and the point was simulated according to U$[0,1]^2$ to describe a small uniform noise effect. Figure \ref{CONTCLU} shows  a realization of the CON (left) and the CLU (right) model.


\begin{table}[ht]
\centering
\begin{tabular}{c|c|cc||c|c|cc}
Alt. & $n$ & $DB$ & $MS$ & Alt. & $n$ & $DB$ & $MS$\\ \hline
\multirow{ 3}{*}{CON} & 50 &  31 & 6 & \multirow{ 3}{*}{CLU} & 50  & 44 & 67\\
& 100 & 58  & 14 & & 100 &  44 & 85\\
& 200 & 89 & 24 & & 200 &   44 & 94 \\ \hline
\end{tabular}
\caption{Empirical rejection rates of DB and MS, unit square}\label{results.unitcube}
\end{table}

Table \ref{results.unitcube} shows the percentages (out of 10000 replications) of rejections of $H_0$ of the distance to boundary test and the
maximal spacing test, rounded to the nearest integer. Obviously, the latter test is sensitive to a cluster alternative, but much inferior to the distance to boundary test
against the contamination alternative.

\renewcommand{\arraystretch}{.7}
\begin{table}[ht]
\centering

\begin{tabular}{c|c|c|ccccccccccccc}
Alt. & $\alpha$ & $n\backslash J$ & 1 & 2 & 3 & 4 & 5 & 6 & 7 & 8 & 9 & 10 & 15 & 20 & 25 \\ \hline
\multirow{ 3}{*}{CON} & \multirow{ 6}{*}{0.5} & 50 & 14 &	22 &	29 &	36 &	43 &	48 &	53 &	56 &	59 &	61 & {\bf 66} & 65 & 60	\\
& & 100 & 19 &	27 &	36 &	45 &	53 &	60 &	66 &	71 &	76 & 	79 & 90 & 93 &{\bf  94} \\
& & 200 & 25 &	38 &	50 &	60 &	68 &	75 &	81 &	86 &	 89 &	91 & 98 & 99 & \last \\ \cline{3-16}
\multirow{ 3}{*}{CLU} &  & 50 & \last & $\ast$ & $\ast$ &  $\ast$ & $\ast$ & $\ast$ &  $\ast$ & $\ast$ & $\ast$ & $\ast$ & $\ast$ & $\ast$ & $\ast$	 \\
& & 100 & \last & $\ast$ & $\ast$ &  $\ast$ & $\ast$ & $\ast$ &  $\ast$ & $\ast$ & $\ast$ & $\ast$ & $\ast$ & $\ast$ & $\ast$  \\ & & 200 & \last & $\ast$ & $\ast$ &  $\ast$ & $\ast$ & $\ast$ &  $\ast$ & $\ast$ & $\ast$ & $\ast$ & $\ast$ & $\ast$ & $\ast$ \\ \hline \hline
\multirow{ 3}{*}{CON} & \multirow{ 6}{*}{2} & 50 & {\bf 11} & {\bf 11} & 9 & 6 & 4 & 3 &	2 & 2 &	1 &	1 & 0 &0 & 1	\\
& & 100 & 21&	27	& {\bf 29}	& {\bf 29}	&26	&23	&19	&14	&10&	7	& 1 & 0 & 0 \\
& & 200 & 33	&50	&59&	65	&68	& {\bf 70} &	{\bf 70}	&69	&68	&66	& 44 & 14 & 1
\\ \cline{3-16}
\multirow{ 3}{*}{CLU} &  & 50 & 39	& {\bf 40}	& 36	& 29	& 22&	16&	12&	8&	6&	5 & 9 & 12 & 13	 \\
& & 100 & {\bf 54} &	52 &	43 &	33 &	24 &	17&	12&	8&	5&	4	& 6 & 10 & 13
 \\
& & 200 & {\bf 64} &	59 &	46 &	32 &	22 &	14 &	9 &	5 &	3 &	2	& 3 & 6 & 8
\\ \hline\hline
\multirow{ 3}{*}{CON} & \multirow{ 6}{*}{5} & 50 & 13 &	18 &	19	& {\bf 20}	& {\bf 20}	& 19	& 17	& 15 &	14 &	12 & 10	& 11 & 13\\
& & 100 & 23 &	34 &	44 &	51 &	56 &	59 &	61 &	{\bf 63} &	{\bf 63} &{\bf 	63} & 53 & 35 & 20\\
& & 200 & 36 &	58 &	74 &	83 &	89 &	92 &	94 &  96 &	97 &	97 & {\bf 98} & 98 & 98
\\ \cline{3-16}
\multirow{ 3}{*}{CLU} &  & 50 & 54	& 63 & {\bf 65} & {\bf 65} & 63 & 61 & 59 & 57 & 55 & 61 & 64 & 61 & 56\\
& & 100 & 78 & 87 & {\bf 88} & 86 & 85 & 83	& 80 & 77 & 74 & 76 & 83 & 81 & 81\\
& & 200 & 93 &	{\bf 98} &	97 &	97 &	96 &	95 &	93 &	91 &	89 &	89 & 94 & 92 & 92 \\ \hline
\end{tabular}
\caption{Empirical rejection rates of the test based on $T_{n,J}^{(\alpha)}$, unit square}\label{results2.unitcube} \end{table}

Table \ref{results2.unitcube} exhibits the corresponding percentages of the test based on $T_{n,J}^{(\alpha)}$. An asterisk denotes power 100\% .
As was to be expected, rejection rates depend crucially on the power $\alpha$ and the the number of neighbors $J$ taken into account.
In each row, the maximum rejection rates have been highlighted using boldface ciphers. The beginning of a sequence of asterisks has also
been emphasized, thus indicating the smallest value of $J$ for which the maximum power is attained. A comparison with Table \ref{results.unitcube}
shows the choice $\alpha = 0.5$ yields a very strong test against cluster alternatives, even for $J=1$. Likewise, taking $\alpha =5$ and any $J \le 15$, the test
based on $T_{n,J}^{(\alpha)}$ outperforms both $DB$ and $MS$.

\subsection{The circle $\mathcal{S}^1$}
\setcounter{equation}{0}
A good overview of tests for uniformity on the circle is presented in the monograph \cite{JS}. We considered the following classical procedures.
\begin{itemize}
\item [(i)] The modified Rayleigh Test, suggested in \cite{Ju2} and denoted by $Ra$ in what follows, is based on the statistic
  \begin{equation}\label{Rayleigh}
  Ra_n:=\left(1-\frac{1}{2n}+\frac{T_n}{8n}\right)T_n.
 \end{equation}
 Here, $T_n:=2n\,|\overline{X}_n|^2$, and $\overline{X}_n = n^{-1}\sum_{j=1}^n X_j$ is the sample mean vector. Under $H_0$, the limit distribution of $Ra_n$ as $n \to \infty$ is the
 $\chi^2_3$-distribution.

\item[(ii)] Kuiper's test (see \cite{Ku}), denoted by $Ku$, uses a transformation of $X_1,\ldots,X_n$ to normed radial data $U_1,\ldots,U_n$, as described in \cite{JS}, p. 153.
 Writing  $0\le U_{(1)}\le \cdots \le U_{(n)}\le 1$ for the order statistics of $U_1,\ldots, U_n$, Kuiper's test is a Kolmogorov-Smirnov type test using the statistic
    \begin{equation*}
    Ku_n:= \sqrt{n}\left(\max_{1\le j\le n}\left(U_{(j)}-\frac{j-1}{n}\right)+\max_{1\le j\le n}\left(\frac{j}{n}-U_{(j)}\right)\right)
    \end{equation*}
    (see \cite{JS}, p. 153).

\item[(iii)] Using the same radial data transformation as in (ii), Watson's test (see \cite{Wat}), denoted by $Wa$, employs the statistic (see \cite{JS}, p. 156)
    \begin{equation*}
    Wa_n:= \sum_{j=1}^n\left(U_{(j)}-\frac{2j-1}{2n} - \frac{1}n\sum_{\ell=1}^nU_\ell + \frac12\right)^2+\frac1{12n}.
    \end{equation*}
    \end{itemize}
The implementation and critical values of the Kuiper (ii) and the Watson (iii) test were taken from the ${\tt R}$-package {\tt Directional}, as provided by \cite{TAS}. As alternative distributions on the circle we considered the von Mises-Fisher (MF) and the Bimodal von Mises-Fisher (BMF) distributions, see \cite{JS}, Section 2.3 and \cite{MJ}, Section 9.3. Note that a unit random vector has the $(d-1)$-dimensional von Mises-Fisher distribution if its probability density function with respect to the uniform distribution is
\begin{equation}\label{vMFdens}
f_{\mu,\kappa}(x)=\left(\frac{\kappa}{2}\right)^{\frac d2-1}\frac1{\Gamma\left(\frac{d}{2}\right)I_{\frac{d}{2}-1}(\kappa)}\exp\left(\kappa\mu'x\right),\quad |x|=1.
\end{equation}
Here, $\kappa > 0$ is a concentration parameter, the unit vector $\mu$ denotes the mean direction, $I_\nu$ is the modified Bessel function of the first kind and order $\nu$, and the prime stands for tranpose. For the simulations in Tables \ref{results.circle} and \ref{results2.circle} we chose $\mu=(1,0)'$ and $\kappa=0.5$. The Bimodal von Mises-Fisher distribution is a mixture of a von Mises-Fisher distribution with $\mu=(1,0)'$ and $\mu=(-1,0)'$ with the same concentration parameter $\kappa=1$.

\begin{table}[ht]
\centering
\begin{tabular}{c|c|ccc||c|c|ccc}
Alt. & $n$ &  $Ra$ & $Ku$ & $Wa$ & Alt. & $n$ &  $Ra$ & $Ku$ & $Wa$ \\ \hline
\multirow{ 3}{*}{MF} & 50 & 58 & 53 & 58 & \multirow{ 3}{*}{BMF} & 50  & 6 & 63 & 61 \\
& 100 & 88& 84& 88& & 100 &  6 & 97 & 99\\
& 200 & $\ast$& 99& $\ast$& & 200 & 6 & $\ast$ & $\ast$\\ \hline
\end{tabular}
\caption{Empirical rejection rates of the tests based on $Ra$, $Ku$ and $Wa$, circle}\label{results.circle}
\end{table}

A comparison of Table \ref{results.circle} and Table \ref{results2.circle} shows that, among the values of $\alpha$ taken into account, the choice $\alpha = 0.05$ and $J=25$ yields the highest power of the new tests against the von Mises-Fisher distribution. This power is comparable with that of $Ra$, $Ku$ and $Wa$ if $n=50$, but the latter tests are superior if $n=100$ or $n=200$. Against the bimodal von Mises-Fisher distribution, the choice $\alpha = 0.5$ and $J=20$ results in a test that outperforms $Ku$ and $Wa$ for $n=50$ and is at least as powerful as these tests if $n=100$ or $n=200$. Against this alternative, the Rayleigh test is not competitive.

\begin{table}
\centering
\renewcommand{\arraystretch}{.7}
\begin{tabular}{c|c|c|ccccccccccccc}
Alt. & $\alpha$ & $n\backslash J$ & 1 & 2 & 3 & 4 & 5 & 6 & 7 & 8 & 9 & 10 & 15 & 20 & 25 \\ \hline
\multirow{ 3}{*}{MF} & \multirow{ 6}{*}{0.5} & 50 & 8 & 	9 &	10 &	12 &	13 &	15 &	16 &	18 &	19 &	21 & 32 & 42 & {\bf 50}	\\
& & 100 & 10 &	12 &	13 &	15 &	17 &	19 &	20 &	21 &	23 &	25	& 34 & 43 & {\bf 53}
 \\
& & 200 & 12 &	15 &	18 &	20 &	23 &	25 &	27 &	30 &	31 &  33 & 43 & 54& {\bf 63} \\ \cline{3-16}
\multirow{ 3}{*}{BMF} & & 50 & 28 &	44 &	56 &	67 &	76 &	82 &	87 &	90 &	93 &	95 & {\bf 98} & {\bf 98} & 97 \\
& & 100 & 37 & 56 &	71 & 80 &	87 &	92 &	95 & 	97 &	98 &	99 & \last & $\ast$& $\ast$ \\
& & 200 & 55 & 77 &	88 & 94 &   97 &	98 &	99 &	\last  & $\ast$& $\ast$& $\ast$& $\ast$& $\ast$\\

\hline	

\multirow{ 3}{*}{MF} & \multirow{ 6}{*}{2} & 50 & 17 &	24 & 29 & 32 &	34 &  {\bf 36} & {\bf 36} &	35 &	33 &	31 & 11 & 1 & 0	\\
& & 100 & 22 & 	33 &	42 &	48 &	53 &	57 &	60 &	63 & 64 & 66	& {\bf 67} & 64 & 50
 \\
& & 200 & 30	& 46 &	58 &	66	& 72	& 77 &	80 &	83	& 85	& 87 	& 92 & 94 & {\bf 95}
\\\cline{3-16}
\multirow{ 3}{*}{BMF} & & 50 & 36 & {\bf 41} &	37 & 	28 & 	16 &	8 &	3 & 	1 & 0 & 0 & 0 & 0 & 0	\\
& & 100 & 64 & 79 &	{\bf 83}  & {\bf 83} &	81 &	77 &	71 &	63 &	53 &	42 & 3 & 0 & 0\\
& & 200 & 86 &  96 &{\bf	99} & {\bf 99} &	{\bf 99} &	{\bf 99} &	{\bf 99}  &	{\bf 99} &	{\bf 99} &	{\bf 99} & 94 & 73 & 28 \\

\hline

\multirow{ 3}{*}{MF} & \multirow{ 6}{*}{5} & 50 & 19 &  27 &	33 &	37 & 	41 &	44 &	46 &	48 &	50 & 	{\bf 51} & 50 & 39 & 15 \\
& & 100 & 24 & 36 &	46 &	52 &	59 &	63  & 67 & 70 &  72 & 75 & 82 & 85 & {\bf 86} \\
& & 200 & 32 & 50 &	62  & 71 & 78 & 83 & 86 & 89 & 91 & 92 & 96 & 97 & {\bf 98} \\
\cline{3-16}
\multirow{ 3}{*}{BMF.} & & 50 & 48 & 61 &	{\bf 66} &	{\bf 66} &	61 & 52 &	41 &	28 &	16 &	8 & 2 & 6 & 18 	\\
& & 100 & 75 & 91 &	96 &	98 &	{\bf 99} &	{\bf 99} &	{\bf 99} &	{\bf 99} &	{\bf 99} &	98 & 92 & 51 & 4 \\
& & 200 & 92 & 99 &	\last  & $\ast$& $\ast$  & $\ast$ & $\ast$  & $\ast$  & $\ast$ & $\ast$ & $\ast$ & $\ast$ & $\ast$\\
\hline
\end{tabular}
\caption{Empirical rejection rates of the test based on $T_{n,J}^{(\alpha)}$, circle}\label{results2.circle}
\end{table}

\subsection{Sphere $\mathcal{S}^2$}
We now treat the case of testing for uniformity on a sphere in $\R^3$, for which many tests have been proposed.
A good overview, also for the corresponding testing problems in higher dimensions,
is given in  \cite{FLE,MJ}. We considered the following procedures.
\begin{itemize}
\item [(i)] The Rayleigh Test (see \cite{Ju2}), denoted by $\widetilde{Ra}$, rejects the hypothesis of uniformity for large values of
 \begin{equation*}
  \widetilde{Ra}_n:=\left(1-\frac{1}{2n}+\frac{T_n}{16n}\right)T_n,
 \end{equation*}
 where $T_n:=2n\,|\overline{X}_n|^2$. Under $H_0$, the limit distribution of $\widetilde{Ra}_n$ as $n \to \infty$ is $\chi^2_3$.

\item[(ii)] The data-driven Sobolev test for uniformity applied to the sphere, here called the Jupp test and denoted by  $JT$ (see \cite{Ju}), computes
\begin{equation*}
  B_n(k)= S_n(k) - k(k+2)\, \log n,
 \end{equation*}
 where
 \begin{equation*}
   S_n(k) = \frac{2k+1}{n}\sum_{j,\ell=1}^nP_k(X_j'X_\ell),
 \end{equation*}
 and $P_k$ is the Legendre polynomial of order $k$. The test statistic is then $JT_n := S_n(\widehat{k})$, where
 \begin{equation}\label{optJupp}
   \widehat{k}= \widehat{k}(n) = \inf\Big{\{}k\in\N: B_n(k)=\sup_{m\in\N}B_n(m)\Big{\}}.
 \end{equation}
 As suggested in \cite{Ju}, p. 1250, a suitable approximation of the supremum in (\ref{optJupp})
 can be done by considering $\sup_{1\le m\le 5}B_n(m)$ instead. Critical values may be obtained from
 the $\chi^2_3$-distribution, since
 $
   JT_n \tod \chi^2_3
 $
 as $n \to \infty$ under the hypothesis $H_0$ of  uniformity.
\end{itemize}
As alternatives we considered the von Mises-Fisher distribution as in (\ref{vMFdens}) with concentration parameter $\kappa =0.5$ and  mean direction to $\mu =(1,0,0)'$.
A second alternative is the  Kent distribution, see \cite{MJ}, p. 176, with density
\begin{equation*}
f_{\mu,\kappa,\beta}(x)=\frac1{c(\beta,\kappa)}\exp\left(\kappa\mu'x+\beta x'(\tau_1\tau_1'-\tau_2\tau_2')x\right),\quad |x|=1.
\end{equation*}
Here, $c(\beta,\kappa)$ is a normalizing constant, and $\tau_1,\tau_2$ and $\mu$ are mutually orthogonal vectors. The references to the Kent distribution in Tables \ref{results.sphere} and \ref{results2.sphere} use
 $\kappa=0.25, \mu=(1,0,0)'$ and $\beta=2$.

The results of the simulation study are given in Tables \ref{results.sphere} and \ref{results2.sphere}. The presented procedure is competitive for sample sizes $n\ge100$ to the other tests and outperforms the modified Rayleigh test for the Kent distribution.
\begin{table}[ht]
\centering
\begin{tabular}{c|c|cc||c|c|cc}
Alt. & $n$ & $\widetilde{Ra}$ & $JT$ & Alt. & $n$ & $\widetilde{Ra}$ & $JT$\\ \hline
\multirow{ 3}{*}{MF} & 50 & 36 & 36 & \multirow{ 3}{*}{Kent} & 50 & 10 & 99\\
& 100 & 66  & 66 & & 100 &  13 & $\ast$\\
& 200 & 94& 94& & 200 &  22 & $\ast$\\ \hline
\end{tabular}
\caption{Empirical rejection rates of $\widetilde{Ra}$ and $JT$, sphere.}\label{results.sphere}
\end{table}

\begin{table}
\renewcommand{\arraystretch}{.7}
\centering
\begin{tabular}{c|c|c|ccccccccccccc}
Alt. & $\alpha$ & $n\backslash J$ & 1 & 2 & 3 & 4 & 5 & 6 & 7 & 8 & 9 & 10 & 15 & 20 & 25 \\ \hline
\multirow{ 3}{*}{MF.} & \multirow{ 6}{*}{0.5} & 50 & 7	& 8 &	9 &	10 &	11 &	13 &	14 &	15 &	16 &	17 & 23 & 27 & {\bf 31}	\\
& & 100 & 8	&9 &	10 &	11 &	12 &	13 &	15 &	16 &	17 &	18	& 25 &32 & {\bf 39}
 \\
& & 200 & 9	& 11& 13 &	14 &	16 &	17  &	19 &	20 &	22 &	24 & 31 & 39 &	 {\bf 47}\\
\cline{3-16}
\multirow{ 3}{*}{Kent} & & 50 & 66 &	87 &	95 &	98 &	99 &	99 &	 \last &  $\ast$  &  $\ast$  &  $\ast$  &  $\ast$ & $\ast$ & $\ast$	\\
& & 100 & 83 & 97 & 99 &  \last  &  $\ast$  &  $\ast$  &  $\ast$  &  $\ast$  & $\ast$  &  $\ast$  &  $\ast$ & $\ast$ & $\ast$\\
& & 200 & 96 &  \last  &  $\ast$  &  $\ast$  &  $\ast$  &  $\ast$  & $\ast$  &  $\ast$  &  $\ast$  &  $\ast$  &  $\ast$ & $\ast$ & $\ast$\\

\hline	

\multirow{ 3}{*}{MF.} & \multirow{ 6}{*}{2} & 50 & 10 &	12 &	{\bf 13} & {\bf 13} &	{\bf 13}  &	{\bf 13} &	12 &	11 &	10 &	8 &	3 & 2 & 1	\\
& & 100 & 13 &	17 & 19 &	22 &	24 &	25 &	{\bf 26} &   {\bf 26} &	25 &	25 & 20 & 14 & 6  \\
& & 200 & 17 & 24 &	30 &	36 &	40 &	43 &	46 &	48 &	50 &	51	 & {\bf 55} & 54 & 50

\\
\cline{3-16}
\multirow{ 3}{*}{Kent} & & 50 & {\bf 14}  & 9 &	4 &	1 & 0 & 0 & 0 & 0 & 0 & 0 & 0 & 0&0	\\
& & 100 & 41 & {\bf 42} & 36 & 27 & 18 &	10 &	5 & 2 & 0 & 0 & 0 &0 & 0\\
& & 200 & 77 & 87 &	{\bf 88} & 86 & 83 &	78 &	71 & 62	& 51 & 41 & 4 & 0& 0\\

\hline
\multirow{ 3}{*}{MF.} & \multirow{ 6}{*}{5} & 50 & 12 &	15 &	18 & 21 & 23 & 25 &	26 & 27 & {\bf 27} & 27 & 26 & 20 & 11 	\\
& & 100 & 15 & 22 &	28 & 33 & 38 & 42 &	44 &	48 &	50 & 51 & 57 & {\bf 59} & {\bf 59} \\
& & 200 & 19 & 31 & 41 &	50 & 57 &   63  &  67  &  71  &  74  &  76  & 84 & 87 & {\bf 90}   \\
\cline{3-16}
\multirow{ 3}{*}{Kent} & & 50 & 29 & {\bf 35} & 34 & 30 &	25 &	18 &	12 &	7 &	4 &	2 & 2 & 8 & 29	\\
& & 100 & 63 & 79 &	86 &	{\bf 88} &{\bf 88} &	87 &	86 &	83 &	80 & 	76 & 40  & 1 & 0 \\
& & 200 & 91 & 99 &  \last  &  $\ast$  &  $\ast$  &  $\ast$  &  $\ast$  &  $\ast$  &  $\ast$  &  $\ast$  &  $\ast$  & 97 & 96 \\

\hline
\end{tabular}
\caption{Empirical rejection rates of the test based on $T_{n,J}^{(\alpha)}$, sphere}\label{results2.sphere}
\end{table}

\section{Conclusions and open problems}
We have introduced a new, flexible class of universally consistent goodness-of-fit tests based on sums of powers of volumes of weighted
$k$th nearest neighbor balls. Under fixed alternatives, scaled versions of the test statistics converge to the $\alpha$-entropy between
probability distributions. The approach is fairly general, since it covers both goodness-of-fit testing for distributions with a compact, 'full-dimensional' support in $\R^d$,
but also on lower-dimensional manifolds embedded in $\R^d$. Our approach requires $J$, the maximum number of neighbors taken into account, to remain fixed as
$n \to \infty$. It would be desirable to obtain limit theorems also for the case that $J=J(n)$ tends to infinity with the sample size $n$.
Another problem is to generalize the theory to cover testing for a parametric family $\{f(\cdot;\vartheta): \vartheta \in \Theta\}$ of densities.
This could be done by substituting
$f(X_i;\widehat{\vartheta}_n)$ for the weight $f_0(X_i)$, where $\widehat{\vartheta}_n$ is a suitable estimator of $\vartheta$, based on $X_1,\ldots,X_n$.

\begin{appendix}

\begin{sidewaystable}[ht]
\centering

\begin{tabular}{c|r|rrrrrrrrrrr}
  $\alpha$ & $n \backslash J$ & 1 & 2 & 3 & 4 & 5 & 6 & 7 & 8 & 9 & 10 & 15 \\
  \hline
  \multirow{ 3}{*}{$0.5^\ast$} & 50 & 0.78 & 2.02 & 3.58 & 5.43 & 7.52 & 9.82 & 12.32 & 15.01 & 17.88 & 20.91 & 38.24 \\
  & 100 & 0.81 & 2.07 & 3.67 & 5.54 & 7.66 & 9.99 & 12.52 & 15.23 & 18.12 & 21.18 & 38.67 \\
  & 200 & 0.83 & 2.12 & 3.73 & 5.62 & 7.76 & 10.11 & 12.66 & 15.40 & 18.31 & 21.38 & 38.97 \\
  \hline
  \multirow{ 3}{*}{$1.5$} & 50 & 1.74 & 5.59 & 12.10 & 21.70 & 34.70 & 51.48 & 72.29 & 97.46 & 127.19 & 161.76 & 414.95 \\
  & 100 & 1.63 & 5.33 & 11.67 & 21.04 & 33.80 & 50.30 & 70.84 & 95.67 & 125.04 & 159.20 & 409.94 \\
  & 200 & 1.54 & 5.14 & 11.33 & 20.53 & 33.11 & 49.37 & 69.63 & 94.19 & 123.27 & 157.12 & 405.75 \\
  \hline
  \multirow{ 3}{*}{$2$} & 50 & 2.90 & 10.18 & 24.03 & 46.44 & 79.42 & 124.97 & 185.22 & 262.12 & 357.71 & 473.86 & 1435.79 \\
  & 100 & 2.65 & 9.62 & 23.06 & 45.02 & 77.52 & 122.63 & 182.27 & 258.48 & 353.32 & 468.77 & 1425.60 \\
  & 200 & 2.47 & 9.17 & 22.25 & 43.72 & 75.65 & 120.05 & 178.89 & 254.30 & 348.28 & 462.78 & 1413.86 \\
  \hline
  \multirow{ 3}{*}{$3$} & 50 & 11.03 & 44.74 & 121.28 & 266.69 & 511.32 & 891.65 & 1449.73 & 2233.59 & 3295.24 & 4692.49 & 19245.32 \\
  & 100 & 9.94 & 41.83 & 116.04 & 258.27 & 501.19 & 882.49 & 1445.57 & 2240.49 & 3324.52 & 4754.26 & 19656.36 \\
  & 200 & 8.87 & 38.72 & 109.44 & 246.81 & 482.75 & 855.10 & 1409.72 & 2193.28 & 3264.21 & 4685.00 & 19603.93 \\
  \hline
  \multirow{ 3}{*}{$4$} & 50 & 54.72 & 257.57 & 788.16 & 1930.49 & 4070.50 & 7751.35 & 13650.51 & 22653.06 & 35839.35 & 54392.65 & 291880.50 \\
  & 100 & 50.79 & 242.73 & 758.58 & 1880.32 & 4024.20 & 7785.46 & 13870.01 & 23246.64 & 37053.41 & 56712.60 & 311282.60 \\
  & 200 & 45.04 & 222.23 & 704.17 & 1776.53 & 3848.90 & 7489.95 & 13461.61 & 22687.94 & 36382.98 & 55935.54 & 312608.60 \\
  \hline
  \multirow{ 3}{*}{$5$} & 50 & 312.65 & 1734.80 & 6015.97 & 16205.09 & 37517.20 & 77431.40 & 147116.35 & 260944.95 & 439900.80 & 707981.50 & 4871049.50 \\
  & 100 & 306.50 & 1698.35 & 5908.46 & 16242.55 & 37930.64 & 79553.08 & 153602.30 & 276095.80 & 471859.40 & 768763.70 & 5476554.00 \\
  & 200 & 281.05 & 1570.43 & 5554.40 & 15449.12 & 36579.58 & 77266.56 & 149773.42 & 271403.80 & 465433.00 & 763477.00 & 5557572.00 \\
  \hline
\end{tabular}%

\caption{Empirical $95\%$ ($\ast$: $5\%$) quantiles of $n^{-1}T_{n,J}^{(\alpha)}$, unit square}\label{crit1}

\end{sidewaystable}

\begin{sidewaystable}[ht]
\centering
\begin{tabular}{c|r|rrrrrrrrrrr}
  $\alpha$ & $n \backslash J$ & 1 & 2 & 3 & 4 & 5 & 6 & 7 & 8 & 9 & 10 & 15 \\
  \hline
\multirow{ 3}{*}{$0.5^\ast$}& 50 & 0.79 & 2.03 & 3.60 & 5.45 & 7.54 & 9.84 & 12.33 & 15.01 & 17.86 & 20.86 & 37.98 \\
  & 100 & 0.81 & 2.08 & 3.68 & 5.56 & 7.68 & 10.02 & 12.55 & 15.27 & 18.16 & 21.21 & 38.66 \\
  & 200 & 0.84 & 2.12 & 3.74 & 5.64 & 7.77 & 10.13 & 12.69 & 15.42 & 18.34 & 21.42 & 39.01 \\
  \hline
  \multirow{ 3}{*}{$1.5$}& 50 & 1.68 & 5.38 & 11.69 & 20.98 & 33.58 & 49.76 & 69.79 & 93.90 & 122.32 & 155.15 & 390.82 \\
  & 200 & 1.58 & 5.21 & 11.42 & 20.63 & 33.18 & 49.41 & 69.59 & 93.96 & 122.79 & 156.32 & 401.12 \\
  & 100 & 1.51 & 5.06 & 11.18 & 20.28 & 32.72 & 48.84 & 68.92 & 93.23 & 122.03 & 155.58 & 401.79 \\
  \hline
  \multirow{ 3}{*}{$2$}& 50 & 2.76 & 9.75 & 23.12 & 44.77 & 76.69 & 120.49 & 178.17 & 251.15 & 341.35 & 450.18 & 1319.84 \\
  & 100 & 2.57 & 9.36 & 22.56 & 44.17 & 76.19 & 120.63 & 179.35 & 254.37 & 347.56 & 460.84 & 1391.83 \\
  & 200 & 2.41 & 9.02 & 21.93 & 43.20 & 74.85 & 118.92 & 177.36 & 252.23 & 345.50 & 459.05 & 1401.71 \\
  \hline
  \multirow{ 3}{*}{$3$}& 50 & 10.60 & 43.34 & 118.42 & 261.50 & 501.48 & 873.17 & 1414.22 & 2164.48 & 3171.86 & 4477.84 & 17233.30 \\
  & 100 & 9.58 & 41.12 & 114.99 & 257.73 & 501.28 & 884.43 & 1447.28 & 2241.43 & 3320.23 & 4740.78 & 19363.40 \\
  & 200 & 8.70 & 38.47 & 109.41 & 247.74 & 485.79 & 861.66 & 1420.44 & 2211.10 & 3291.23 & 4721.69 & 19694.42 \\
  \hline
  \multirow{ 3}{*}{$4$}& 50 & 52.85 & 250.52 & 773.08 & 1900.67 & 4007.25 & 7624.74 & 13366.61 & 21984.31 & 34446.25 & 51695.70 & 254726.46 \\
  & 100 & 49.10 & 239.95 & 762.58 & 1905.83 & 4108.24 & 7956.09 & 14162.64 & 23708.80 & 37708.27 & 57628.95 & 309717.34 \\
  & 200 & 44.64 & 222.63 & 715.20 & 1821.47 & 3969.98 & 7720.87 & 13872.85 & 23382.36 & 37446.71 & 57560.69 & 318814.82 \\
  \hline
  \multirow{ 3}{*}{$5$}& 50 & 298.41 & 1675.10 & 5838.66 & 15867.96 & 36735.05 & 75663.47 & 142943.40 & 250659.04 & 417322.93 & 662873.89 & 4100114.35 \\
  & 100 & 297.07 & 1704.63 & 6065.70 & 16774.09 & 39537.66 & 83076.29 & 160282.91 & 287664.48 & 489264.08 & 791912.14 & 5477284.59 \\
  & 200 & 278.18 & 1595.17 & 5727.94 & 15994.55 & 38089.13 & 80827.49 & 156752.23 & 284457.16 & 487537.46 & 797292.10 & 5757879.69 \\
   \hline
\end{tabular}
\caption{Empirical $95\%$ ($\ast$: $5\%$) quantiles of $n^{-1}T_{n,J}^{(\alpha)}$, circle.}\label{crit2}
\end{sidewaystable}

\begin{sidewaystable}[ht]
\centering
\begin{tabular}{c|r|rrrrrrrrrrr}
  $\alpha$ & $n \backslash J$ & 1 & 2 & 3 & 4 & 5 & 6 & 7 & 8 & 9 & 10 & 15 \\
  \hline
  \multirow{ 3}{*}{$0.5^\ast$}& 50 & 0.78 & 2.01 & 3.58 & 5.43 & 7.51 & 9.82 & 12.33 & 15.02 & 17.88 & 20.92 & 38.28 \\
  & 100 & 0.81 & 2.07 & 3.67 & 5.54 & 7.66 & 9.99 & 12.52 & 15.24 & 18.13 & 21.19 & 38.68 \\
  & 200 & 0.83 & 2.12 & 3.73 & 5.62 & 7.76 & 10.11 & 12.66 & 15.40 & 18.31 & 21.39 & 38.97 \\
  \hline
  \multirow{ 3}{*}{$1.5$}& 50 & 1.74 & 5.57 & 12.07 & 21.64 & 34.62 & 51.35 & 72.12 & 97.19 & 126.83 & 161.28 & 413.52 \\
  & 100 & 1.63 & 5.33 & 11.66 & 21.04 & 33.81 & 50.29 & 70.78 & 95.60 & 124.97 & 159.14 & 409.60 \\
  & 200 & 1.54 & 5.14 & 11.34 & 20.54 & 33.11 & 49.37 & 69.63 & 94.16 & 123.21 & 157.03 & 405.72 \\
  \hline
  \multirow{ 3}{*}{$2$}& 50 & 2.89 & 10.15 & 23.98 & 46.38 & 79.42 & 125.09 & 185.38 & 262.19 & 357.62 & 473.44 & 1429.69 \\
  & 100 & 2.65 & 9.63 & 23.07 & 45.00 & 77.51 & 122.55 & 182.25 & 258.53 & 353.50 & 469.04 & 1425.90 \\
  & 200 & 2.46 & 9.18 & 22.24 & 43.71 & 75.61 & 119.96 & 178.79 & 254.13 & 348.09 & 462.50 & 1413.42 \\
  \hline
  \multirow{ 3}{*}{$3$}& 50 & 11.05 & 44.70 & 121.62 & 267.35 & 514.39 & 899.85 & 1465.00 & 2259.71 & 3334.17 & 4753.49 & 19354.84 \\
  & 100 & 9.88 & 41.61 & 115.67 & 258.08 & 501.29 & 882.86 & 1446.89 & 2243.22 & 3329.77 & 4766.78 & 19759.49 \\
  & 200 & 8.89 & 38.78 & 109.53 & 247.04 & 483.45 & 856.25 & 1409.95 & 2195.31 & 3266.71 & 4688.33 & 19641.06 \\
  \hline
  \multirow{ 3}{*}{$4$}& 50 & 55.17 & 257.13 & 789.21 & 1934.64 & 4099.03 & 7841.76 & 13896.74 & 23139.96 & 36671.41 & 55854.21 & 298772.54 \\
  & 100 & 50.44 & 241.42 & 756.21 & 1882.42 & 4046.14 & 7822.84 & 13949.15 & 23382.01 & 37302.88 & 57241.48 & 315602.34 \\
  & 200 & 45.06 & 221.64 & 706.06 & 1778.94 & 3853.76 & 7503.19 & 13470.18 & 22738.34 & 36479.18 & 56091.39 & 313949.22 \\
  \hline
  \multirow{ 3}{*}{$5$}& 50 & 312.86 & 1751.03 & 6086.85 & 16525.35 & 38243.32 & 79307.62 & 151325.83 & 270378.36 & 456890.66 & 738839.85 & 5085970.95 \\
  & 100 & 304.65 & 1690.91 & 5915.53 & 16290.68 & 38389.23 & 80548.51 & 155432.36 & 279993.35 & 478289.61 & 780194.53 & 5581455.18 \\
  & 200 & 285.05 & 1578.28 & 5582.39 & 15450.00 & 36521.11 & 77144.17 & 149839.80 & 272028.17 & 466294.26 & 763822.88 & 5578854.38 \\
\hline
\end{tabular}
\caption{Empirical $95\%$ ($\ast$: $5\%$) quantiles of $n^{-1}T_{n,J}^{(\alpha)}$,  sphere.}\label{crit3}
\end{sidewaystable}

\end{appendix}

B. Ebner and N.Henze, Institute of Stochastics, Karlsruhe Institute of Technology (KIT), Englerstr. 2, D-76133 Karlsruhe:
\\
{\texttt Bruno.Ebner@kit.edu} \ \ {\texttt Norbert.Henze@kit.edu}
\vspace*{2mm}


J. E. Yukich, Department of Mathematics, Lehigh University,
Bethlehem PA 18015:
\\
{\texttt joseph.yukich@lehigh.edu}

\end{document}